# Models of Vortices and Spirals in White Dwarf's Accretion Binaries

Daniela Boneva

*Space and Solar Terrestrial Research Institute – Bulgarian Academy of Sciences, 6 Moskovska St, 1000 Sofia, Bulgaria*
*danvasan@space.bas.bg*

**Abstract.** The main aim in the current survey is to suggest models of the development of structures, such as vortices and spirals, in accretion white dwarf's binaries. Numerical methods and simulations are applied on the bases of hydrodynamic analytical considerations. It is suggested in the theoretical model the perturbation's parameters of the accretion flow, which are caused by the influences of the tidal wave over the flux of accretion matter around the secondary star. The results of numerical code application on the disturbed flow reveal an appearing of structure with spiral shape due to the tidal interaction in the close binaries. Our further simulations give the solution, which expresses the formation of vortical configurations in the accretion disc's zone. The evolution of vortices in areas of the flow's interaction is explored using single vortex and composite vortex models. Gas in the disc matter is considered to be compressible and non-ideal. The longevity of all these structures is different and each depends on the time period of the rotation, density and velocity of the accretion matter.

## INTRODUCTION

One of the energy sources in binary stars systems with white dwarfs is the accretion process. The vortices and spirals are known as the efficient mechanisms of angular momentum transportation, which are studied by many authors. Petersen et al. [1], Klahr and Bodenhiemer [2] investigated the growth of vortices in disc via the baroclinicity conditions, using numerical methods. Johnson and Gammie [3] have performed series of runs with zero initial vorticity and perturbation wavelengths and give one possibility of what generates it. Sawada, Matsuda and Hachisu have presented in [4], by using numerical methods, the appearing of spiral shock waves, as tidally induced shocks. Spiral structures in white dwarfs are detected by Steeghs et al. [5]. The observational proofs of the existence of structures are based on the indirectly methods, such as the Doppler Tomography, developed by Marsh in [6]. Our results of tomography are reviewed in the Discussion chapter.

## Description of Set of the Equations and Methods

The character of interaction between the flow of matter and envelop of two star's components brings to the necessity of applying the equations of gasdynamics. We are using the well-known gas-dynamical equations, which are given previously by many authors [7], [8]: equation of mass conservation; Navier-Stokes equations, which takes a principal role in our calculations and given by Thorn in [9]; energy balance equation for a viscous non-ideal fluid, equation of the angular momentum conservation and equation of state for compressible flow. We obtained the expression, known as the vortical transport equation the way it was generated by Nauta in [10].

The applying of numerical methods is necessary for the calculations herewith and they are in accordance with mathematical software, supporting the examinations here. We apply the method of Runge-Kutta in the combination with "Alternating direction implicit method" (ADI), [11] and [12]. Free boundary conditions were considered on the outer disc edge, where the density is defined to be constant: $\rho_{out} = 10^{-8} \rho_{L_1}$, where $\rho_{L_1}$ is the density of the inner Lagrangian point ~ $L_1$. We adopted the spherical form of the place of interaction, with a free radius. The most appropriate theoretical method to examine the transitional states of stability or instability in the considering system is based on bifurcation analysis [13].

## Results

The changes in mass transfer rate and interaction of the flows in the binary may cause disturbances in the flow parameters. We study the perturbation quantities of velocity and density, as we applied perturbation function on the Navier-Stokes equations. The results are published in [14].

### *Vorticity undulations in the accretion disc area*

We use the baroclinicity character of the accretion flow, which is in the relation of non-conservancy form of the vortical transport equation. As a result, we present the simulation of vortex kind of pattern growth in plane of the disc zone. The picture's frames are visualizing a covered range of about $7.687 \times 10^{-6}$ AU to $6.68 \times 10^{-5}$ AU that is the boundary area of performing the calculation.

The development of such vortices passes through three stages, following the runs of calculations. First, a distortion of the luminary flow is observed. In the next series, the layers in the examined area undergo weekly undulation. Finally, the values of density and velocity during the last calculation time range confine to the stage of the vortex evolution in some steady period of their development, when they are "ready" for the angular momentum transportation. Similar results were published in [15].

### *Formation of spiral-like structures*

We are studying tidally generated spiral waves that are the result of interaction of the inflow stream, coming from the donor with the matter of the accretor.

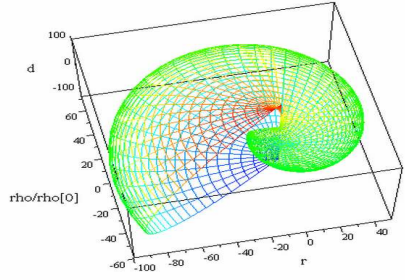

**FIGURE 1.** Spiral's formations in the disc flow area. The figure shows the development of one spiral arm, when the gas is hot in two stages of spiral's development.

The gas in the disc is considered to be hot and then the solutions show appearing of one-arm spiral-like wave structure. As seen in Fig. 1 that this wave may be penetrated into the inner parts of the disc's configuration. The figures show spiral's formations, as the separated from the whole disc images. This way, the form of the wave is more decipherable, Fig.1.

### *Structure's formation as a result of the Turing bifurcation method*

According to the spatial-temporal characteristic of the exploring formations, applying the method of the so-called Turing bifurcation is the most appropriate. We carry out an expression, which is an analogue of the structure formation and it follows the model of the "reaction-diffusion" type equation of [16]. The right hand-side of the equation gives the combination of the reaction part, confirmed to the established disc's structure and the diffusion part, related to the angular momentum distribution. We carry out the next view of the equation, describing the cause and effect of the structure's changes in the disc:

$$\frac{\partial \Psi_r}{\partial t} = (\nabla . v_r)\Psi_r - (\nabla . \Psi_\varphi)v_\varphi + \frac{D_r}{D_\varphi}\nabla^2 \Psi_r \quad (1a)$$

$$\frac{\partial \Psi_\varphi}{\partial t} = (\nabla . v_r)\Psi_r - (\nabla . \Psi_\varphi)v_\varphi + \frac{D_r}{D_\varphi}\nabla^2 \Psi_\varphi \quad (1b)$$

where: $\Psi$ - is the vorticity; $D$ - is the diffusion coefficient ( or transport coefficient); $v$ - is the velocity of the flow. Analysis of Eq. 1a, 1b gives us the difference in values of the diffusion coefficient components: $D_r \approx 2.75 D_\varphi$. This result points to the necessary condition for the appearing of Turing bifurcation. Further activity of Turing bifurcation may cause an appearing of spherical 3D structures of vortical type.

## CONCLUSION

By using the methods of numerical calculations in a combination with the gas-dynamical equations, bifurcation analysis and observational methods we obtained the results, which give us a physical explanation of the morphology of the accretion gas flow. Tidally interaction of the inflow from donor star with the matter of circumdisc halo causes development of spiral density wave, as one-armed model for the hot disc and of two-dimensional vorticity development in outer disc edges, by applying numerical methods and mechanisms of Turing bifurcation.

Using the technique of Doppler Tomography as the observational method in a combination with the gas-dynamical simulation, we confirmed that the spiral structures in the accretion disc in the white dwarfs close binaries are the main elements of the flow. We have derived Hβ and Hγ Doppler tomograms for SS Cyg in its active state from spectral observations of the star, see papers [17], [18].

## ACKNOWLEGMENTS

We wish to thank the organizers of the 17[th] European White Dwarf Workshop for all support to present and publish this paper.

## REFERENCES

1. M. R. Petersen, G. R. Steward, K. Julien, Ap J., 658, 2007, pp.1252-1263
2. H. Klahr, P. Bodenheimer, ApJ, 582, 2003, pp. 869-892
3. B. M. Johnson, C. F. Gammie, ApJ, 635, 2005, pp. 149-156
4. K. Sawada, T. Matsuda, I. Hachisu, MNRAS, 219, 1986, pp. 75
5. D. Steeghs, E. T. Harlaftis, K. Horne, MNRAS, 290, 1997, L28.
6. T. R. Marsh, MNRAS, 231, 1988, p. 1117
7. J. Frank, A. King, D. Raine, Accretion Power in Astrophysics, 3-rd edition, Cambridge University Press, 2002, New York
8. J. Dyson., D. Williams, Physics of the Interstellar Medium, 2000, Manch. University Press
9. K. Thorne, Foundations of fluid dynamics, 2004, V 0415.2.K2004,
10. M. D. Nauta, Two-dimensional vortices and accretion disks, 2000, University Utrecht
11. K. K. Autar, E. K. Egwu, Numerical methods with applications, 2008, 1 ed., self-publ., http://numericalmethods.eng.usf.edu/topics/textbook_index.html
12. M. J. Chang, L.C. Chow, W.S. Chang, Numerical Heat transfer, Part B, 19(1), 1991, pp. 69-84, ISSN 1040-7790
13. R. Seydel., Practical bifurcation and stability analysis, v.5, Springer, 1994, 3 ed.
14. D. V. Boneva, BgAJ, 11, 2009, pp. 53–65
15. D. V. Boneva, BgAJ, 13, 2010, pp. 3-11
16. R. Engelhardt, Modelling Pattern Formation in Reaction-Diffusion Systems, 1994, Univ. of Copenhagen
17. D. A. Kononov, P. V .Kaigorodov, D. V. Bisikalo, A. A. Boyarchuk, M. I. Agafonov, O. I. Sharova, A.Yu.Sytov, and D. V. Boneva, Astron. Rep., vol. 85, No. 10, 2008, pp. 927-939
18. D. Boneva, P. V. Kaigorodov, D. V. Bisikalo, and D. A. Kononov, Astron. Rep., 2009, Vol. 53, No. 11, 2009, pp. 1004–1012, Pleiades Publishing, Ltd.,. ISSN 1063-7729